\documentclass[11pt,twoside]{article}
\usepackage{asp2006}
\usepackage{epsf}
\usepackage{./psfig}
\usepackage{lscape}
\usepackage[dvips]{graphicx} 
\def\ltsim{~\rlap{\lower-0.5ex\hbox{$<$}}{\lower 0.5ex\hbox{$\sim\,$}}}
\def\mabs{\hbox{$M_{\rm abs}$}\,}
\newcommand{\typeo}{Type~I } \newcommand{\typeoc}{Type~I}
\newcommand{\typet}{Type~II } 
\newcommand{\typeiii}{Type~III } \newcommand{\typeiiic}{Type~III}
\newcommand{\typect}{Type~II-CT } \newcommand{\typectc}{Type~II-CT}
\newcommand{\typeolr}{Type~II.o-OLR }
\newcommand{\typeolrc}{Type~II.o-OLR}
\begin{document}
\title{Three Types of Galaxy Disks} 
%%%
\author{M. Pohlen\altaffilmark{1}, P. Erwin\altaffilmark{2}, I. Trujillo\altaffilmark{3}, and J.E. Beckman\altaffilmark{3}} 
%%%
\affil{\altaffilmark{1}Kapteyn Astronomical Institute,
University of Groningen, NL-9700 AV Groningen, The Netherlands}
\affil{\altaffilmark{2}Max-Planck-Institut f\"ur extraterrestrische
Physik, D-85748 Garching, Germany}
\affil{\altaffilmark{3}Instituto de Astrof\'{\i}sica de Canarias,
E-38200 La Laguna, TF, Spain}
%%% 
%%%
\begin{abstract} 
We present our new scheme for the classification of radial
stellar surface brightness profiles for disk galaxies. We summarize
the current theoretical attempts to understand their origin and give an
example of an application by comparing local galaxies with their
counterparts at high redshift ($z\!\approx\!1$).\\[-0.8cm]
\end{abstract}
%%%
%%%
%%%%%%%%%%%%%%%%%%%%%%%%%%%%%%%%%%%%%%%%%%%%%%%%%%%%%%%%%%%%%%%%%%%%%%%%%
\section{Eclectic Introduction}  
Galaxy formation and evolution is one of the most fascinating topics
in astrophysical research today. Along several main pathways we are
trying to answer the question of how galaxies are formed and how they
evolve over time within a cosmological framework for structure
formation.
Theoretically this is often done via numerical N-body/SPH simulations
\citep[e.g.][]{governato2007} or semi-analytical modelling
\citep[e.g.][]{somerville1999}.
According to \cite{freeman2002} the associated observational part could
be divided into doing {\it near-field cosmology}, i.e.~looking for the fossil
records of the galaxy formation and evolution process by observing
local galaxies in detail, or doing {\it far-field cosmology}, i.e.~looking
at distant objects for the progenitors of modern-day galaxies.
Although it is now observationally possible to obtain spatially
resolved kinematics and emission-line measurements as a function of
radius out to a redshift of $z\!\approx\!1$ \citep{weiner2006},
surface photometry is still a valuable source of information.
By modelling the surface-brightness distribution of a galaxy and
thereby parametrising the individual components, we can obtain a
common ground to measure, compare, and sort large samples of galaxies.
The structures we observe, our fossil records, should be linked to the
galaxy assembly. 
%
%
%%%%%%%%%%%%%%%%%%%%%%%%%%%%%%%%%%%%%%%%%%%%%%%%%%%%%%%%%%%%%%%%%%%%%%%%%
\section{Sample, Data, and Profile Classification}
To create a local reference data set for comparison with high redshift
galaxies we have collected imaging data for two large, complementary
samples of face-on to intermediate inclined disk galaxies.
The first is a diameter and distance limited sample of early-type
galaxies (S0-Sb) split into two subsamples: 66 strong- or weakly
barred galaxies \citep*{erwin2005,erwin2007} and 45 unbarred galaxies
(Aladro et al., this volume) both drawn from the UGC catalog. The
images are obtained with a variety of different telescopes including
the Sloan Digital Sky Survey (SDSS, DR5).
The second sample is a volume limited sample ($D\ltsim 46\,$Mpc) down
to a limiting magnitude ($\mabs\!<\!-18.4\,$B-mag) of late-type
(Sb-Sdm) galaxies by \cite{pohlen2006}. It comprises all selected
galaxies from the LEDA on-line catalog having useful imaging data
available in the SDSS (DR2), the sole data source for this study.
Details about the extraction of the analyzed surface brightness
profiles can be found in the cited papers.
In our analysis we concentrated not only on providing statistics on
the exponential disk as given by \cite{dejong1996} or
\cite{macarthur2004}, but rather study the shape of the profiles, as not
all --indeed only the minority-- are well described with a single
exponential fitting function. 
To do so, we revised and extended the classification of surface
brightness profiles introduced in the pioneering paper by \cite{free70}
including the so called truncation of the stellar population at the
edge of the disk discovered by \cite{vdk1979}.
\begin{figure}[t]
\plotfiddle{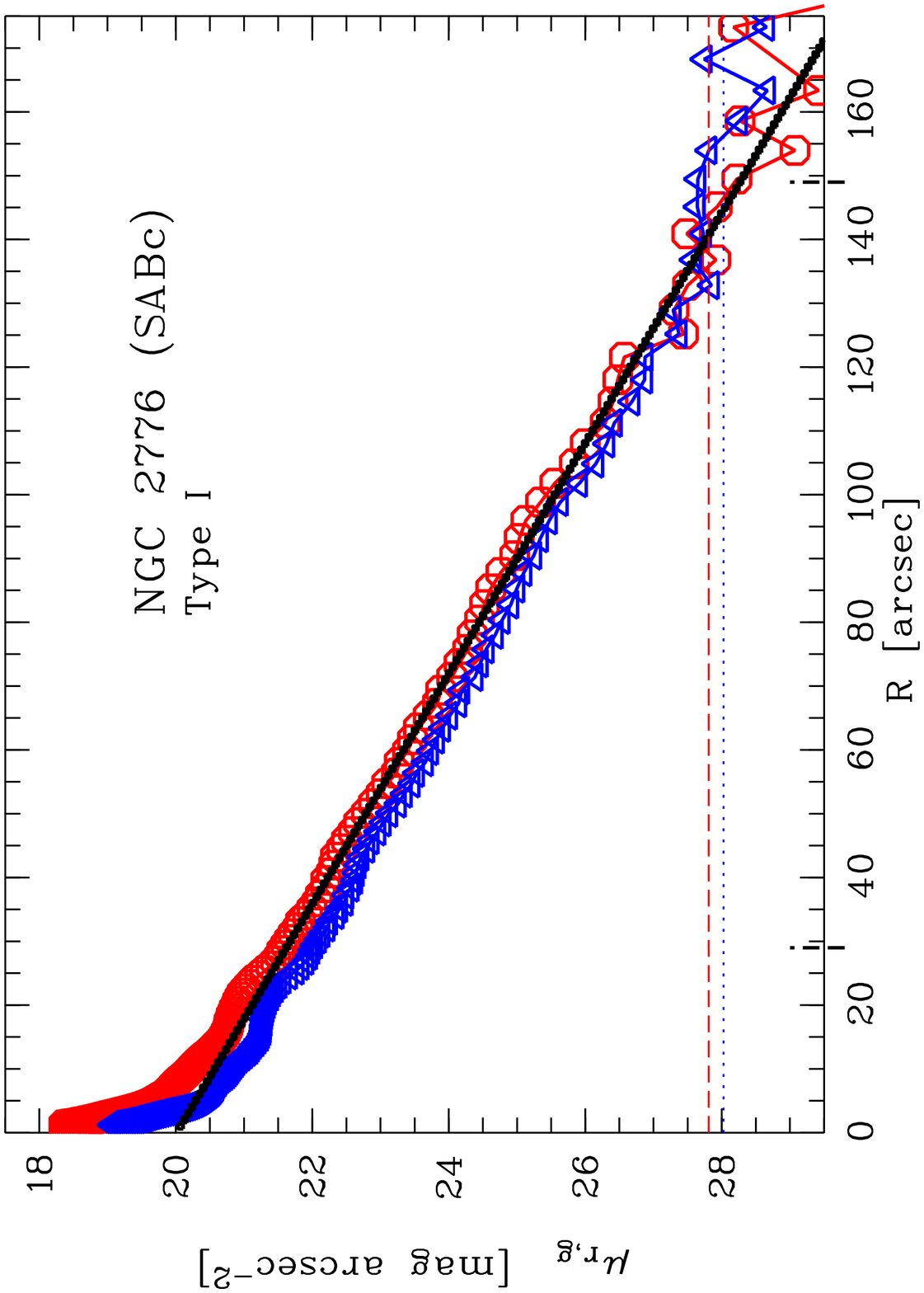}{0cm}{270}{17}{17}{-195}{23} 
\plotfiddle{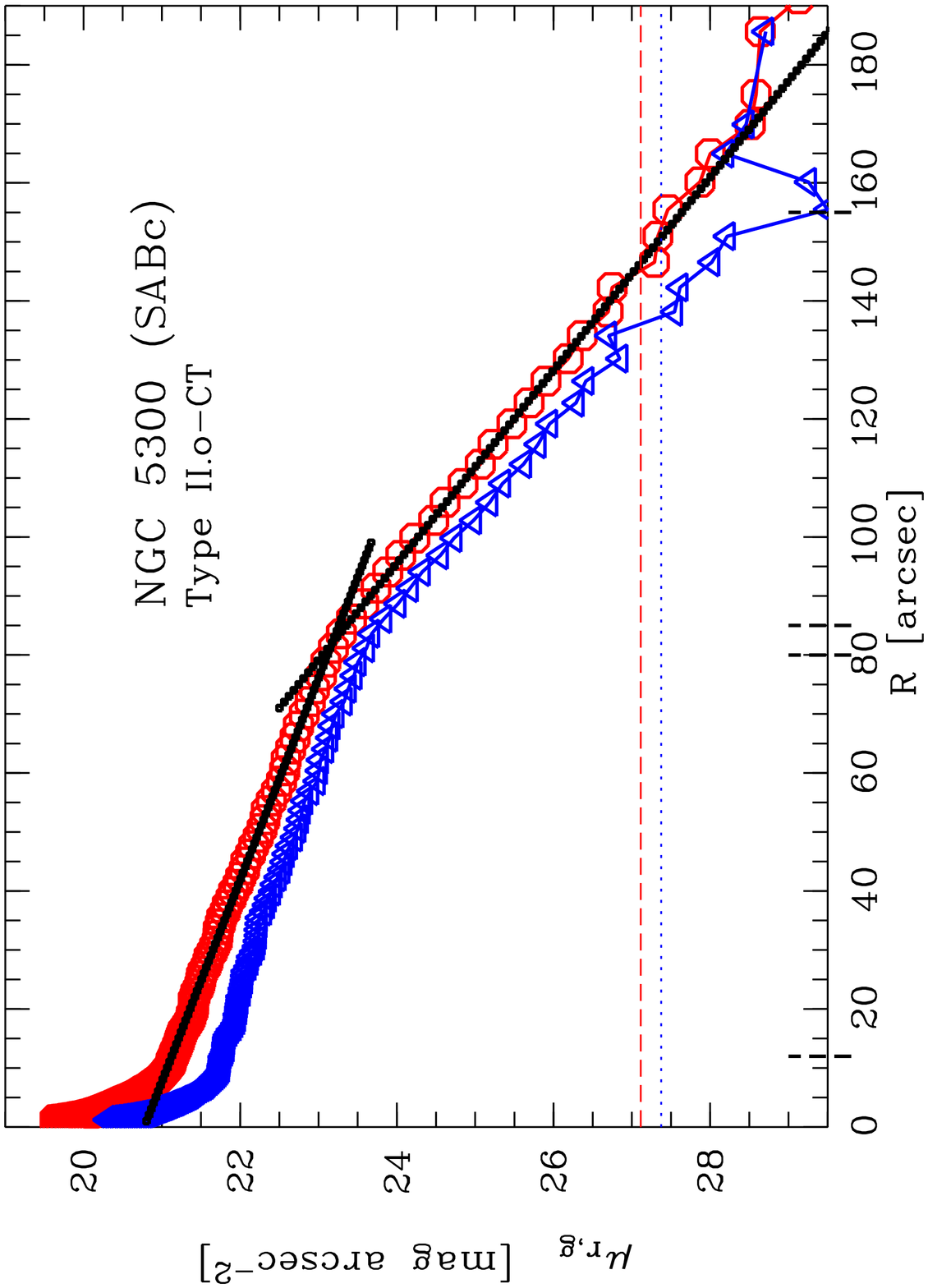}{0cm}{270}{17}{17}{-68}{47.5} 
\plotfiddle{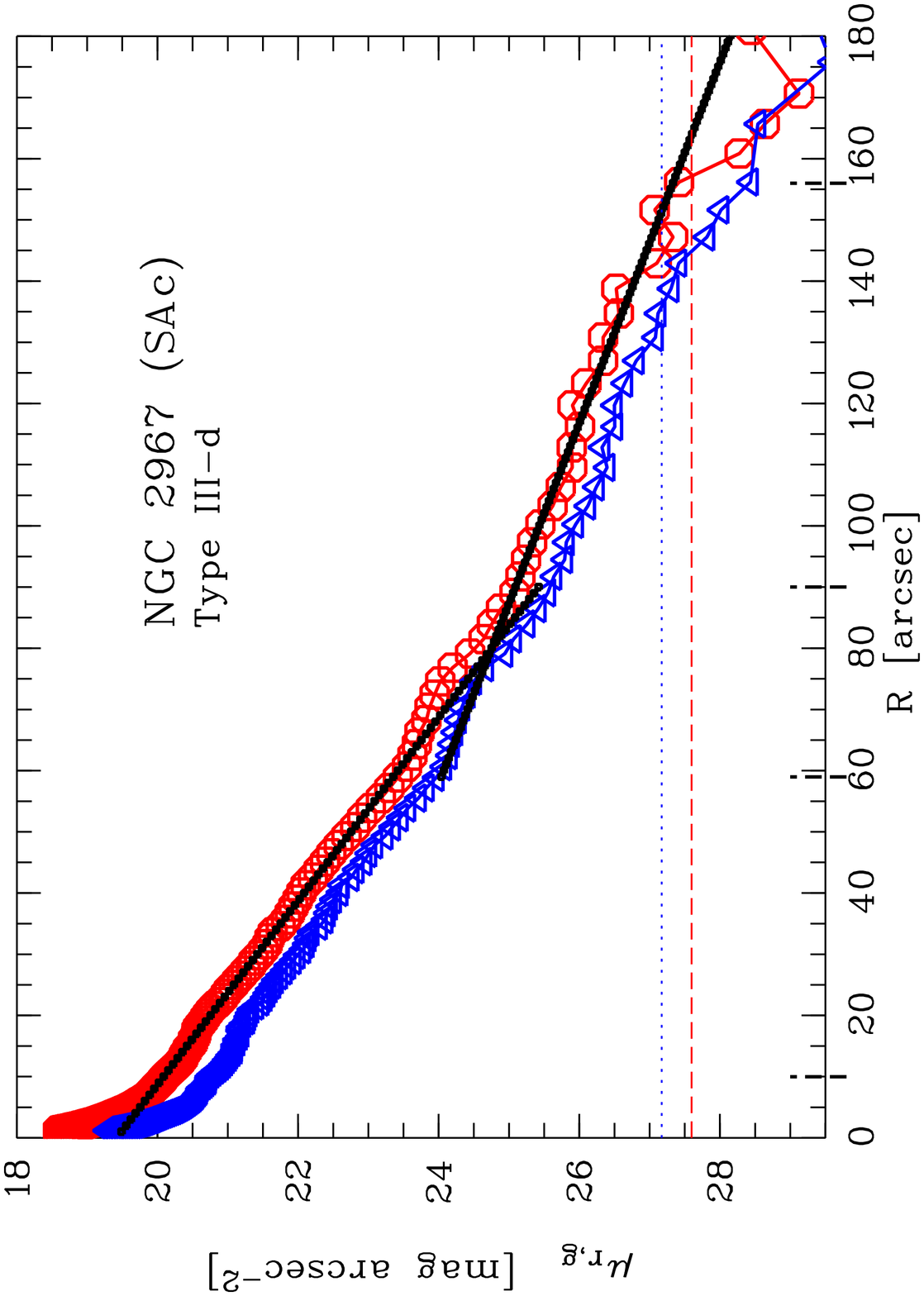}{0cm}{270}{17}{17}{60}{72} 
\vspace*{0.6cm}
\caption{The three main disk types: Type\,I, Type\,II, and Type\,III
{\sl (from left to right)}.  Azimuthally averaged, radial SDSS surface
brightness profiles in the g'- {\it (triangles)} and r'- {\it (circles)}
band overlayed by r'-band exponential fits to the individual regions:
single disk; inner and outer disk. }
\end{figure}
We identified three basic classes of surface brightness profiles
depending on an apparent break feature or lack of one (cf.~Fig.1): 1)
The well known Freeman \typeo that has an exponential profile, with no
break.  2) \typet with a 'downbending break', now including the
truncations.  3) A completely new class, called \typeiiic, also
described by a broken exponential but with an upbending profile,
shallower beyond the break. The latter, discovered by
\cite{erwin2005}, is also termed {\it anitruncated} since showing the
opposite behavior to the well known truncated profiles.
We have taken great care in ensuring that the critical sky subtraction
is done properly, therefore always providing lower levels down to
where we trust our profiles (cf.~Fig.1).  It turns out that the
uncertainty in our sky subtraction, in almost all cases, does not
interfere with the basic classification of a galaxy.
Nevertheless, it is worth noting that there are now completely
independent measurements (from resolved star counts) of three local
counterparts for each of our main types. There is NGC\,300
\citep{Bland-Hawthorn2005}, where an unchanging exponential has been
detected out to as far as ten scalelengths, M\,33 \citep{ferguson2006},
where the profile is well fitted by a broken exponential with a
downbending break, and M\,31 \citep{ibata2005}, which could be
regarded as a \typeiii (antitruncated) disk.

In order to address the origin of the different types we have
introduced an interpretative subclassification scheme for our
profiles. For example, we distinguish between those \typet profiles
possibly associated with a bar resonance, called \typeolrc, and those
probably related to a star-formation threshold, called {\it classical
truncation} or \typect (see next section, as well as
Erwin et al. and Aladro et al., this volume).
Evidence about the origin may come from the trends we observed with
Hubble type (see Fig.2). Any model of galaxy formation and evolution
should be able to explain why we find for example more \typect cases
in late-type than early-type galaxies, whereas the trend is
the other way around for the \typeo galaxies.
\begin{figure}[t]
\plotfiddle{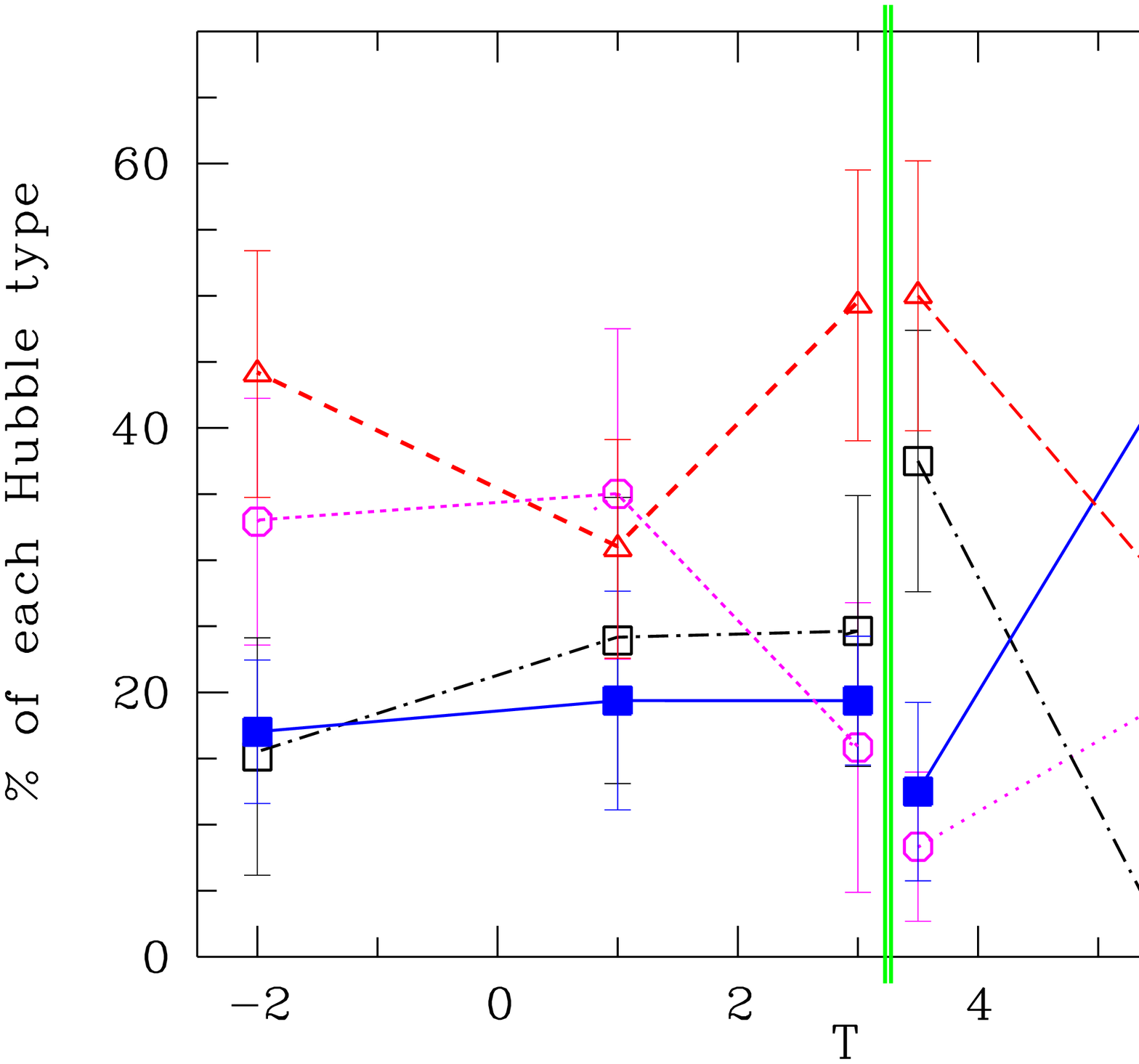}{0cm}{0}{30}{30}{-150}{-150} 
\vspace*{4.7cm}
\caption{ Frequency of profile types \typeo {\it (open circles)},
\typect {\it (filled squares)}, \typeiii {\it (open triangles)}, and
\typeolr {\it (open squares)} in relation to the Hubble type. The
galaxies are merged in six morphological bins ($T\!\in\![-3.4$--$-0.5],[-0.4$--$1.4],[1.5$--$3.4], [2.5$--$4.4], [4.5$--$6.4], [6.5$--$8.4]$). 
The vertical lines
separate the results of the early- and late-type sample.}
\end{figure}
\section{Origin and Implications}
Our basic classification is simply morphological in nature. We do
pure taxonomy similar to grouping galaxies along the Hubble
sequence. But what is the physical background for the different types,
what does theory tell us about surface brightness profiles? 
It turn out our current physical picture lags behind the
observational evidence. The origin of purely exponential disks
(\typeoc) in itself is not fully understood, although well known since
the 1940-50's. One of the most successful attempts to show
that an exponential disk can arise naturally out of disk formation
conditions was by \cite{yos} who showed that an exponential disk will
form if the viscosity time scale and the star formation time scale are
comparable. 
The currently favored model ascribes the classical truncations
(\typectc) to the effect of a star formation threshold in disk column
density as the edge is approached \citep{kennicutt89,schaye2004,
elmegreen2006,li2006}.
On the other hand, using purely collisionless $N$-body simulations
\cite{debattista2006} find downbending breaks appearing in the disk
profile as a resonance phenomenon, which may be associated to our
\typeolr breaks.
Observationally, there is circumstantial evidence that some of the
galaxies with \typeiii profiles are associated with recent mergers
(e.g. M31), supported now by the new N-body/SPH simulations of
\cite{younger2007} showing that minor mergers could produce upbending
stellar profiles in the remnant galaxy. On the other hand, by a small
change of the input gas density \cite{elmegreen2006} are also able to
produce \typeiii profiles within their evolutionary model of an
isolated galaxy with star formation.
Once the nature of the galactic stellar disk is determined it will be
fully justified to use the shape of the profiles and the position
of the breaks for comparison of galaxies at various redshifts as done
by \cite{perez2004}, \cite{trujillo2005} or \cite{tamm2006}.
For example, \cite{trujillo2005} presented a first direct application
of the new classification scheme by exploring the Hubble Ultra Deep
Field. 
Using the position of the truncation as a direct estimator of the size
of the stellar disk it becomes possible to observe inside-out growth
of galactic disks --inherent in the current galaxy formation and
evolution models-- comparing the local late-type \typet galaxies
with their counterparts at high-z (i.e.~$z\!\approx\!1$) in a
straightforward manner.
Although still uncertain because of the small number of galaxies
analyzed, their results suggest that the radial position of the
truncation has increased with cosmic time by $\sim\,1\!-\!3$ kpc in the
last $\sim\,8$\,Gyr indicating a small to moderate ($\sim\!25\%$)
inside-out growth of the disk galaxies since $z\!\sim\,1$.
%
%%%%%%%%%%%%%%%%%%%%%%%%%%%%%%%%%%%%%%%%%%%%%%%%%%%%%%%%%%%%%%%%%%%%%%555
%
\acknowledgements %%% Text of acknowledgements runs on after this command.
M.P.~would like to thank the organisers and the Kapteyn Instituut for
their generous financial support. P.E.~was supported by DFG Priority
Program 1177. This work was supported by grant AYA2004.08251-CO2-01 of
the Spanish Ministry of Education and Science, and project P3/86 of
the IAC. \\[-0.7cm]
%
%%% THE BIBLIOGRAPHY
%%%


\begin{thebibliography}{}
\bibitem[Bland-Hawthorn et al.(2005)]{Bland-Hawthorn2005} Bland-Hawthorn,
J., Vlaji{\'c}, M., Freeman, K.~C., \& Draine, B.~T.\ 2005, ApJ, 629, 239
%
\bibitem[Debattista et al.(2006)]{debattista2006} Debattista, V.~P., 
Mayer, L., Carollo, C.~M., Moore, B., Wadsley, J., \& Quinn, T.\ 2006, 
ApJ, 645, 209 
%
\bibitem[de Jong(1996)]{dejong1996} de Jong, R.~S. 1996, A\&A 313, 45
%
\bibitem[Elmegreen \& Hunter(2006)]{elmegreen2006} Elmegreen, B.~G., 
\& Hunter, D.~A.\ 2006, ApJ, 636, 712 
%
\bibitem[Erwin et al.(2005)Erwin, Beckman \& Pohlen]{erwin2005} Erwin, P., 
Beckman, J.E., \& Pohlen, M.\ 2005, ApJL, 626, 81 
%
\bibitem[Erwin et al.(2007)Erwin, Pohlen \& Beckman]{erwin2007} Erwin, P., 
Pohlen, M., \& Beckman, J.E.\ 2007, AJ, sub.
%
\bibitem[Ferguson et al.(2006)]{ferguson2006} Ferguson, A., 
Irwin, M., Chapman, S., Ibata, R., Lewis, G., \& Tanvir, 
N.\ 2006, astro-ph/0601121 
%
\bibitem[Freeman(1970)]{free70} Freeman K.C. 1970, ApJ, 160, 811
%
\bibitem[Freeman \& Bland-Hawthorn(2002)]{freeman2002} Freeman, 
K., \& Bland-Hawthorn, J.\ 2002, ARA\&A, 40, 487 
%
\bibitem[Governato et al.(2007)]{governato2007} Governato, F., et al.\
2007, MNRAS, 374, 1479
%
\bibitem[Ibata et al.(2005)]{ibata2005} Ibata, R., Chapman, S., 
Ferguson, A.~M.~N., Lewis, G., Irwin, M., \& Tanvir, N.\ 2005, ApJ, 634, 
287 
%
\bibitem[Kennicutt(1989)]{kennicutt89} Kennicutt, R.C., 1989, ApJ, 344, 685
%
\bibitem[Li et al.(2006)Li, Mac Low \& Klessen]{li2006} Li, Y., Mac
Low, M.-M., \& Klessen, R.~S.\ 2006, ApJ, 639, 879
%
\bibitem[MacArthur et al.(2004)]{macarthur2004} MacArthur, L.~A., 
Courteau, S., Bell, E., \& Holtzman, J.~A.\ 2004, \apjs, 152, 175 
%
\bibitem[P{\'e}rez(2004)]{perez2004} P{\'e}rez, I.\ 2004, 
A\&A, 427, L17 
%
\bibitem[Pohlen \& Trujillo(2006)]{pohlen2006} Pohlen, M., 
\& Trujillo, I.\ 2006, A\&A, 454, 759  
%
\bibitem[Schaye(2004)]{schaye2004} Schaye, J. 2004, ApJ 609, 669 
%
\bibitem[Somerville \& Primack(1999)]{somerville1999} Somerville, 
R.~S., \& Primack, J.~R.\ 1999, \mnras, 310, 1087 
%
\bibitem[Tamm \& Tenjes(2006)]{tamm2006} Tamm, A., \& Tenjes, 
P.\ 2006, A\&A, 449, 67 
%
\bibitem[Trujillo \& Pohlen(2005)]{trujillo2005} Trujillo, I., 
\& Pohlen, M.\ 2005, ApJL, 630, L17 
%
\bibitem[van der Kruit(1979)]{vdk1979} van der Kruit, P.~C., 1979,  
A\&AS 38, 15
%
\bibitem[Weiner et al.(2006)]{weiner2006} Weiner, B.~J., et al.\ 
2006, \apj, 653, 1027 
%
\bibitem[Yoshii \& Sommer-Larson(1989)]{yos} Yoshii Y., \& Sommer-Larson J. 1989, MNRAS 236, 779
%
\bibitem[Younger et al.(2007)]{younger2007} Younger, J.D.,  T. J. Cox,T.J., Anil C. Seth,A.C., \& Hernquist, L., 2007, ApJ, sub. 
%
%
\end{thebibliography}
\end{document}